# Universal behavior of the apparent fragility and the structural relaxation time for the high pressure previtreous effect

Aleksandra Drozd-Rzoska

Institute of High Pressure Physics Polish Academy of Sciences,

ul. Sokołowska 29/37, 01-142 Warsaw, Poland

**ABSTRACT**

This report shows the universal previtreous behavior of the pressure related apparent fragility, i.e. the steepness index $m_T(P) = d\,log_{10}\,\tau(P)/d(P/P_g)$, where the glass transition temperature and pressure are defined as $\tau(T_g, P_g) = 100s$. It is shown that $m_T(P) = 1/(P^* - P)$, for $P \to P_g$, with the singular pressure $P^* = P_g - \Delta P^*$. Basing on this finding, the 3-parameter relation for portraying the previtreous behavior of the primary (structural) relaxation time is derived: $\tau(P) = \tau_0^P (P^* - P)^{-\Phi}$. The fair portrayal of experimental data was shown for glass forming 8*OCB (liquid crystal), EPON 828 (resin) and diisobutyl phthalate, propylene carbonate (low molecular weight liquids). On decompressing, at $P_B \sim (P_g - 0.5GPa)$, the dynamic crossover from the 'close-to-$P_g$' to the 'remote-from-$P_g$' dynamic domains is clearly detected. It is in fair agreement with the 'magic' crossover time-scale $\tau(P_B) \sim 10^{-7}s$. All these recalls the mode-coupling-theory behavior, but for the pressure path, and the possible general universality of the dynamic crossover time scale within the $\tau(T, P)$ space. It is notable, the discussed description offers the fair portrayal of experimental data even in the extreme pressure range $0.1 MPa < P < 2.2 GPa$.

I. INTRODUCTION

Glass transition constitutes the grand challenge of the solid state physics, chemical physics, geophysics and many branches of material engineering. However, despite more and more advanced experimental and theoretical insights, the long-awaited cognitive breakthrough still seems to be distant [1-5]. From the experimental side, one of promising strategies that can bring the long-awaited cognitive breakthrough closer is the focus on specific experimental approaches. This can be the wider implementation of *in situ* high-pressure research, which developed intensively in the last decades decades ago [4, 6-14]. The fundamental importance of such studies is associated with the fact that the shift of pressure affects mainly density and the free volume whereas basic temperature studies are coupled to the activation energy, with a (very) minor impact of density / free volume changes. This basic difference offers a unique possibility of decoupling many phenomena overlapping in studies under atmospheric pressure [15]. Some specific properties observed under atmospheric pressure may appear the consequence of a phenomenon 'hidden' within the P-T plane, for instance a continuous phase transition [16]. Coherent pressure and temperature studies of dynamic properties matched with the P-V-T (pressure – volume – temperature) enabled also the direct insight into the volume path, what led to the discover of the so called thermodynamic scaling [13, 17]. Notwithstanding, the problem of most basic and characteristic feature of the glass transition, namely the previtreous behavior of the primary relaxation time (or alternatively viscosity), remains puzzling, particularly for the pressure path [2-6]. Generally, it is expected that in the previtreous domain above the glass pressure $P_g$ [4, 6]:

$$\tau(P), \eta(P) \propto exp\left(\frac{PV_a(P)}{RT}\right) = exp(PV_a^{'}(P)) \qquad (1)$$

where $T = const$ and $P < P_g$, and $V_a^{'}(P) = V_a(P)/RT$ is the normalized apparent, pressure dependent, activation volume.

At the end of the 19th century, Barus [18] considered the pressure dependence of viscosity in liquids and suggested the description via the relation $\eta(P) \propto exp(AP)$, $A = const$. Hence, eq. (1) can be named as the super-Barus (SB), by the analogy to the super-Arrhenius (SA) equation with apparent activation energy $E_a(T)$: $\tau(T), \tau(T) \propto exp(E_a(T)/RT)$, $P = const$ used in temperature studies under atmospheric pressure [4]. Neither the SA nor SB relation cannot be applied directly for describing experimental data due to unknown general forms of $E_a(T)$ or $V_a(T)$ and then 'ersatz' relations have to be used. In 1972 Johari and Whalley [19] measured viscosity on approaching the glass transition in supercooled glycerol, and portrayed obtained experimental data via:

$$\tau(P) = \tau_0 \, exp\left(\frac{B}{P_0 - P}\right) \tag{2}$$

where $P > P_g$ and $T = const$; the extrapolated singular pressure $P_0 > P_g$ and the amplitude $B = const$.

The parallel relation was implemented for describing the pressure evolution of the primary relaxation time studies in a set of glass formers in 1999 [20]. Since the behaviour of $\eta(P)$ and $\tau(P)$ are parallel, the discussion will be focused on the latter. The careful analysis of the $\tau(P)$ experimental data obtained from the high resolution broad band dielectric spectroscopy (BDS) showed the (eq. (2)) generally fails for so called 'fragile' glass formers, i.e. strongly different from the basic Barus dependence [21]. 1n 1996 the following dependence was proposed to overcome this problem [21]:

$$\tau(P) = \tau_{ref.} \exp\left(\frac{D_P P}{P_0 - P}\right), \tag{3}$$

where $D_P$ is the fragility strength coefficient for the pressure path. The comparison of Eqs. (2) and (6) yields $V_a(T) = RTD_P/(P_0 - P)$. The prefactor $\tau_{ref.}$ is determined by the pressure evolution of the relaxation time under atmospheric pressure, hence it can range between $10s$ and $10^{-10}s$ for different isotherms.

It is notable that eq. (3) can be reduced to the basic Barus equation, what is not possible for eq. (2). In subsequent decades eq. (3) has become the basic tool for portraying the previtreous behevior in super-pressed glass forming systems [4-6]. It is notable that eq. (3) contains parameter $D_P$, directly related to the fragility. Since its appearance it is one of basic concepts of the glass transition physics, being the metric enabling the 'universal' describing of the previtreous dynamics for microscopically different glass formers [2-8]. It has been introduced by Angell et al. [22, 23] when considering the plot $log_{10}\eta(T)/(T_g/T)$ and $log_{10}\tau(T)/(T_g/T)$ vs. $T_g/T$, with the assumption $\tau(T_g, P_g) = 100s$ and $\eta(T_g, P_g) = 10^{13} Poise$. This plot enabled a common presentation of previtreous changes of the primary (structural) relaxation time and viscosity for different low molecular weight liquids and polymers. The fragility coefficient was defined as $m = [d\,log_{10}\eta(T)/d(T_g/T)]_{T=T_g} = [d\,log\,\tau(T)/d(T_g/T)]_{T=T_g}$ [22, 23]. For the basic Arrhenius behaviour ($E_a(T) = const$) one obtains $m = m_{min} = log_{10}\tau(T_g) - log_{10}\tau_0 = 16$, assuming the 'quasi-universal' value for the prefactor in the SA equation $\tau_0 = 10^{-14}s$ [4, 22, 23]. Glass formers with dynamics relatively close to the basic Arrhenius pattern ($m < 30$) are named 'strong' and systems with notable distortions from the basic pattern are called 'fragile' ($m > 30$). By analogy, for the pressure path the fragility was defined as $m_T = d\,log_{10}\tau(P)/d\,log(P/P_g)$ [4, 5, 22, 23]. The index 'T' was introduced to stress the isothermal and pressure related nature.

This report focuses on yet undiscussed issues of the universal previtreous ($P < P_g$) behavior of the apparent pressure fragility and the resulted from this finding new equation for portraying the pressure evolution of the primary (structural) relaxation time or viscosity.

## II. EXPERIMENTAL

This report presents results of high pressure studies, up to $P \sim 2.2$ GPa, of the structural relaxation time in few glass formers: liquid crystalline (8*OCB, isooctyloxycyanobiphenyl) [10], epoxy resin EPON 828 [9, 11] and two low molecular weight liquids diisobutyl phtalate (DIIP) [9, 12] and propylene carbonate [9, 12, 14]. Due to the location and the form of the pressure dependence of the glass temperature $T_g(P)$ studies for 8*OCB and EPON 828 were carried out for two isotherms (and in the range of pressures between the atmospheric one ($P = 0.1 MPa$) and $P \sim 0.5 GPa$ for the time scale from $\tau \sim 10^{-7} s$ do $\tau(T_g) = 100 s$. For DIIP and PC the range of tested pressure extends up to $P \sim 1.25 GPa$ $P \sim 2.21 GPa$, respectively, and the smallest time scale reached $\tau(P = 0.1 MPa) \sim 10^{-10} s$. All these was obtained via the broad band dielectric spectroscopy (BDS) measurements using the Novocontrol BDS analyzer, model 2015. Structural relaxation times were determined from peak frequencies of primary relaxation times loss curves $\varepsilon''(f)$ as $\tau = 1/2\pi f_{peak}$. It is notable that high pressure *in situ* BDS studies for $f > 10 MHz$, related to time scale $\tau > 10^{-6} s$, remain the non-solved experimental challenge. To overcome this problem $\tau(T)$ experimental data were matched and scaled with essentially low frequency DC electric conductivity $\sigma(T)$ experimental data determined via the relation $\sigma = (2\pi f)\varepsilon''(f)$. The scaling of $\tau(T)$ and $\tau_\sigma(T) \propto 1/\sigma(T)$ experimental data was possible due to the fact that for highest frequencies obeys the translational – orientational coupling: $\tau(T)\sigma(T) = const.$ [4] Samples were placed in the module with the flat parallel measurement capacitor, with the gap $d = 0.1 mm$, which was subsequently placed in the high pressure chamber. The measurement voltage $U = 1V$ was used.

## III.     RESULTS AND DISCUSSION

The long-range previtreous behaviour of viscosity or the primary (structural) relaxation time is the hallmark of the glass transition physics [1-6]. Figure 1 shows the pressure evolution of the structural relaxation time for supercooled/ superpressed liquid crystal (8*OCB), epoxy resin (EPON 828) and two low molecular weight liquids: diisobutyl phthalate (DIIB) and propylene carbonate (PC). For 8*OCN and EPON 828 results are for two isotherms. It is notable that presented experimental data extends from the atmospheric pressure ($P = 0.1 MPa$) to the glass transition pressure, determined by the $T_g(P)$ curves. For tested systems they are given in refs. [9, 10, 12, 14]. It is notable that for DIIB $T_g \approx 1.23 GPa$ and for PC $T_g \approx 2.27 GPa$; for 8*OCB and EPON 828: $T_g < 0.54 GPa$. Consequently, results for the two latter compounds are additionally shown in Figure 2 to get better insight into details, particularly regarding the fitting quality.

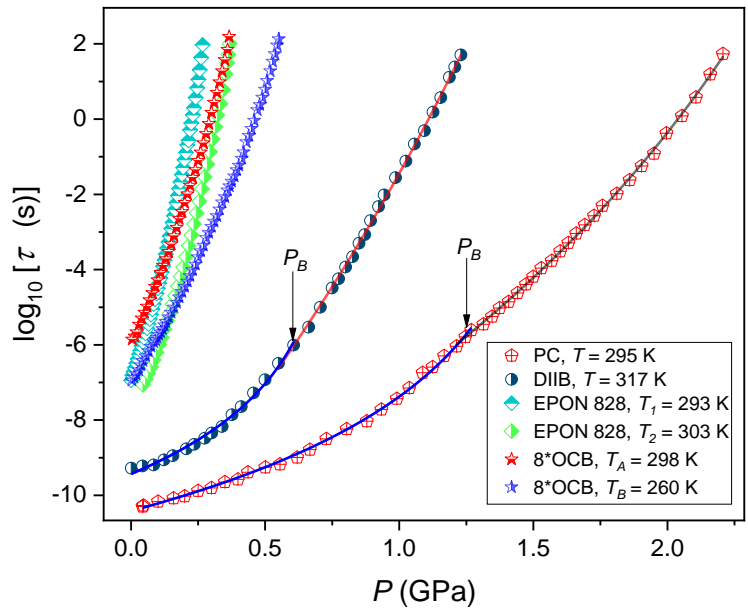

**Figure 1**     Pressure dependences of the primary relaxation time for tested glass forming systems. In each case $\tau(P)$ data are terminated in the immediate vicinity of the glass pressure: $\tau(P_g) \sim 10^2 s$: their values are collected in Table I. Experimental data are portrayed by eq. (7), with parameters also given in Table I. Note the manifestation of the dynamic crossover phenomenon (Arrows and Table I), associated with changes of values of parameters in eq, (7).

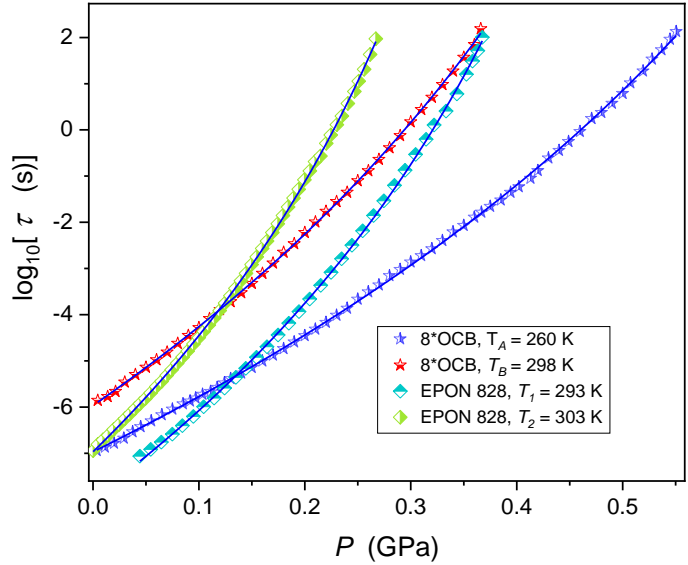

**Figure 2**  Pressure dependences of the primary relaxation time for 8OCB and EPON 828, for two isotherms in each case. Experimental data are portrayed by eq. (7), with parameters given in Table I.

For the broad-pressure-range previtreous effects of the structural relaxation time the apparent fragility (steepness indexes) were calculated via [4, 6]:

$$m_T(P) = \frac{d\,log_{10}\,\tau(P)}{d(P/P_g)} \quad (4)$$

where $0.1 MPa < P < P_g$, $T = const$.

Results of such analysis are presented below, in Figures 3 and 4: separately for 8*OCB, EPON 828 and DIIB, PC , to reach the better insight because of the strongly different pressures domains. For all tested systems, belonging to the group liquid crystals, resins and low molecular weight liquids, the same 'universal' form of the previtreous behaviour of the apparent fragility is visible:

$$\frac{1}{m_T(P)} = a + bT \qquad (5)$$

where the singular pressure $(P^*)$ is easily determined from the conditions $1/m_T(P^*) = 0$.

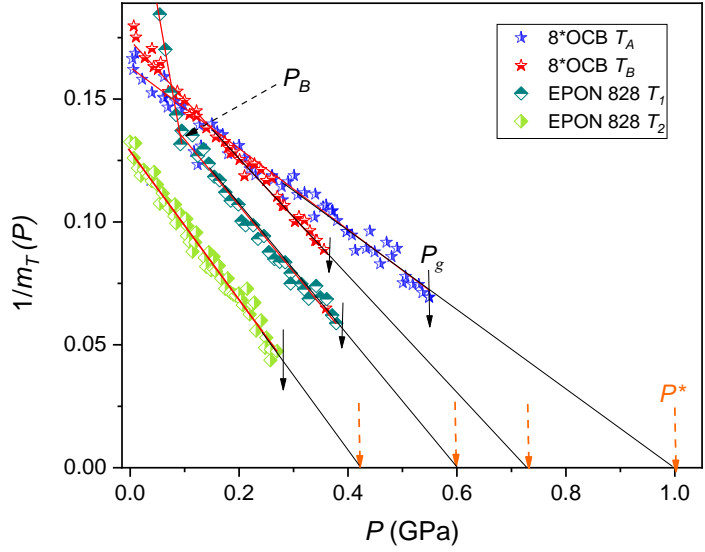

**Figure 3** The universal previtreous behavior of the reciprocal of the pressure apparent fragility for supercooled 8*OCB and EPON 828, basing on experimental data from Figs. 1, 2. Tested isotherms recalls values are also given in these Figures. Results of the linear regression fit (eq. (5)) are collected in Table II (Appendix). The solid arrows indicated examples glass transition pressures, and the dashed (orange) arrows extrapolated singular pressures $P^* = P_g + \Delta P^*$. For the 'high temperature' isotherm for EPON 828 the dynamic crossover pressure $P_B$ is also indicated.

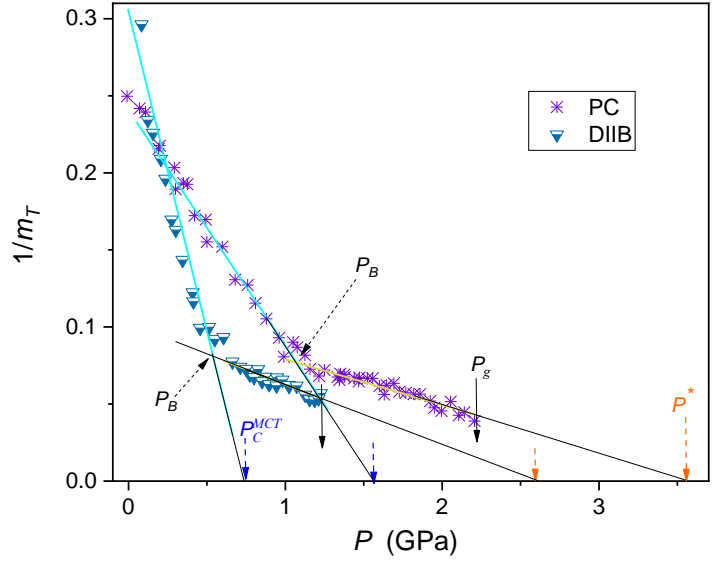

**Figure 4** The universal previtreous behavior of the reciprocal of the pressure apparent fragility for supercooled propylene carbonate and diisobutyl phthalate, basing on experimental data from Figs. 1, 2. Tested isotherms recalls values are also given in these Figures. Results of the linear regression fit (eq. (5)) are collected in Table II (Appendix). The solid arrows indicated glass transition pressures, and the dashed (orange) arrows extrapolated singular pressures $P^* = P_g + \Delta P^*$. Dashed arrows in blue are related to the upper 'remote-from-$P_g$' dynamical domain and domain and then $P^* = P_C^{MCT}$. Dynamical crossover pressures $P_B$ associated with changes of value of parameter in eq, (5), see Table II, are also indicated.

Linking eqs. (5) and (6) one obtains:

$$d\,log_{10}\,\tau(T) = \frac{d(P/P_g)}{aP+b} \qquad (6)$$

The integration integrating yields the following relation for the pressure evolution of the structural relaxation time:

$$\tau(P) = \tau_0^P (P^* - P)^{-\Phi} \qquad (7)$$

for $P < P_g$ and the singular pressure $P^* = P_g + \Delta P^*$

Tables 1 and II in the Appendix presents results of fittings for the apparent fragility (eq. (5), linear regression fit) and for the relaxation time via eq.(7) (nonlinear 3 parameter fit). Result of fitting are graphically presented in Figures 1, 2 and Figures 3, 4 - respectively. Notable is the superior agreement for linear ranges of $[m_T(P)]^{-1}$ vs. $P$ behavior (eq. 5) and domains of the validity of eq. (7). The same situation takes place for singular pressures $P^*$, what suggest that the nonlinear fitting via eq. (7) can be reduced solely to two parameters (prefactor $\tau_o^P$ and the exponent $\Phi$), because values of $P^*$ can be easily estimated from the condition $[m_T(P^*)]^{-1}$. It seems that the analysis $[m_T(P)]^{-1}$ vs. $P$ can also serve as the model – free tool for estimating the dynamic crossover location: it seems to emerge for the 'high temperature' isotherm for EPON 828 ($P_B = 0.138 GPa, \tau_B = 4\times10^{-6}$) and for DIIB ($P_B = 0.544 GPa$, $\tau_B = 2.3\times10^{-7}$) and PC ($P_B = 1.075 GPa$, $\tau_B = 2.5\times10^{-7}$). The superior agreement between the dynamic crossover 'magic' time scale suggested by Novikov and Sokolov [24] from the analysis of $\tau(T)$ experimental data $\tau_B = 10^{-7\pm1} s$ and pressure results takes place. When commenting this results, it is worth stressing that the analysis presented in Figs. 1 and 2 avoids the assumed *a priori* validity of a model equation, used in analysis carried out so far [4, 6, 25].

It is also notable that the critical type behavior of $\tau(P)$ (eq. (7) ), associated with linear domains of $[m_T(P)]^{-1}$ in Figs. 3 and 4, is related to different types of power exponent in the high pressure (close-to-$P_g$: $\Phi = 24 - 55$) and low-pressure (remote-from-$P_g$: $\Phi = 5 - 7$) domains. In the opinion of the author the latter can be associated with the mode-coupling theory (MCT) behavior [4, 26], and then it can be considered as its first clear evidence under high pressures.

## IV. CONCLUSIONS

This report presents the evidence for the presumably 'universal' behavior of the apparent fragility $m_T(T) \propto 1/(P^* - P)$, offering also a simple and reliable estimation of the extrapolated 'spinodal' pressure $P^*$ as well as the 'discontinuity' of the glass transition for the pressure path: $\Delta P^* = P^* - P_g$, $\tau(T_g, P_g) = 100\,s$ or $\eta(T_g, P_g) = 10^{13}\,Poise$. The previtreous 'anomaly' of the apparent fragility served subsequently as the base for the derivation of the relation $\tau(P) = \tau_0^P (P^* - P)^{-\phi}$, for which the fair ability for portraying experimental data was shown in the broad range of pressures. When comparing fittings of experimental data via the new eq. (7) and the 'former' eq. (3) the singular pressure $P^*$ is notably lower than . It seems that future heat capacity and the structural entropy studies under pressure can be decisive here, because they may reveal a possible coincidence with ideal glass [4] 'Kauzmann pressure' and its pressure – related 'dynamic' estimations (eq. (3) or eq. (7)). Worth stressing is the possibility of the simple and model free estimation of the dynamic crossover pressure: the dominated so far methodology assumed *a priori* the validity of the VFT equation [4, 25] or its 'pressure counterpart (eq. (3)) [6] for describing $\tau(T, P)$ experimental data. The discussed previtreous behavior described by the 'critical-like' eq. (7) is well visible also for the 'remote-from-$P_g$' dynamical domain, what can be considered as the first clear evidence for the mode-coupling-theory (MCT) validity also for the pressure path. In summary, this report shows the new and presumably universal previtreous 'anomaly' of the pressure related apparent fragility and based on this finding pressure evolution of the structural relaxations time. Such behavior can be also expected to appear for viscosity, electric conductivity, diffusion, …in the previtreous domain. For obtaining the discussed evidence important was the extreme range of teste pressures, reaching even 2.2 GPa.

**ACKNOWLEDGEMENT**

This research was carried out due to the support of the National Centre for Science (Poland), project NCN OPUS ref.  2016/21/B/ST3/02203, head Aleksandra Drozd-Rzoska.

**APPENDIX.**

**TABLE I** Results of fitting of experimental data $\tau(P \to P_g)$, for the primary (structural) relaxation time via eq. (7).

| Fitted Relation | | $ln\,\tau(P) = ln\,C_P - \Phi \times log(P^* - P)$ | | | |
|---|---|---|---|---|---|
| | | Parameters | | | |
| | | $\Delta P_{range}[GPa]$ | $ln\,C_P[s]$ | $\Phi$ | $P^*[GPa]$ |
| Glass Forming Materials | DIIP | $\Delta P_{remote}$ 0.01 ÷ 0.60<br>$\Delta P_{close}$ 0.56 ÷ 1.24 | -10.00<br>13.53 | 5.44<br>55.20 | 0.78<br>2.66 |
| | PC | $\Delta P_{remote}$ 0.05 ÷ 1.26<br>$\Delta P_{close}$ 1.21 ÷ 2.21 | -8.94<br>4.09 | 7.19<br>29.38 | 1.61<br>3.42 |
| | EPON 828 (1) | 0.05 ÷ 0.38 | -13.30 | 23.70 | 0.60 |
| | EPON 828 (2) | 0.04 ÷ 0.33 | -15.00 | 23.90 | 0.46 |
| | 8*OCB (2) | 0.01 ÷ 0.37 | -9.60 | 26.80 | 0.73 |
| | 8*OCB (1) | 0.01 ÷ 0.55 | -6.80 | 26.50 | 1.02 |
| | | $\Delta P_{range} = |P_r - P_g|$, $P_r$ denotes the terminal pressure<br>$\Delta P_{close}$ - close to $P_g$; $\Delta P_{remote}$ - remote from $P_g$ | | | |

**TABLE II** Results of the linear regression fit for the reciprocal of apparent fragility $m_T(P)$: see eq. (5).

| Fitted Relation | | $\dfrac{1}{m_T(P)} = aP + b$ ; $aP + b \neq 0$ | | | |
|---|---|---|---|---|---|
| | | Parameters | | | |
| | | $\Delta P_{range}[GPa]$ | $a$ | $P^*[GPa]$ | $P_g[GPa]$ |
| | DIIP | $\Delta P_{remote}$ 0.01 ÷ 0.66<br>$\Delta P_{close}$ 0.60 ÷ 1.25 | -0.415<br>-0.039 | (#) 0.73<br>2.60 | 1.23 |
| | PC | $\Delta P_{remote}$ 0.05 ÷ 1.26<br>$\Delta P_{close}$ 1.18 ÷ 2.21 | -0.24<br>-0.03 | (#) 1.56<br>3.50 | 2.22 |
| | EPON 828 (1) | $\Delta P_{remote}$ 0.05 ÷ 0.096<br>$\Delta P_{close}$ 0.086 ÷ 0.39 | -1.25<br>-0.33 | (#) 0.92<br>0.59 | 0.28 |
| | EPON 828 (2) | 0.01 ÷ 0.27 | -0.30 | 0.42 | 0.39 |
| | 8*OCB (2) | 0.01 ÷ 0.55 | -0.24 | 0.73 | 0.55 |
| | 8*OCB (1) | 0.01 ÷ 0.38 | -0.16 | 1.00 | 0.365 |
| | | $\Delta P_{range} = |P_r - P_g|$, $P_r$ denotes the terminal pressure<br>$\Delta P_{close}$ - close to $P_g$; $\Delta P_{remote}$ - remote from $P_g$ ; (#) $P^* = P_C^{MCT}$ | | | |